\documentclass[fleqn,usenatbib]{mnras}

\usepackage{newtxtext,newtxmath}

\usepackage[T1]{fontenc}
\usepackage{ae,aecompl}

\usepackage{graphicx}	
\usepackage{amsmath}	
\usepackage{amssymb}	

\usepackage{rotating, graphicx}
\usepackage{subfig}
\usepackage{pdflscape}
\usepackage{natbib}
\usepackage{amsmath}
\usepackage{courier}
\usepackage{color, colortbl}
\usepackage{float}
\restylefloat{table}
\usepackage{footnote}
\usepackage[T1]{fontenc}
\usepackage{aecompl}
\usepackage{amssymb}
\usepackage{rotating}
\usepackage{mathtools}
\usepackage{enumitem,booktabs,cfr-lm}
\usepackage[referable]{threeparttablex}
\renewlist{tablenotes}{enumerate}{1}
\usepackage{multirow}
\usepackage{ulem}

\def\aap{Astronomy \& Astrophysics}
\def\apj{The Astrophysical Journal}

\def\mnras{Monthly Notices of the Royal Astronomical Society}

\def\apjs{The Astrophysical Journal Supplement}
\def\apjl{ApJL}
\def\nat{Nature}

\title[NGC 300 microquasar jet]{Optical IFU spectroscopy of a bipolar microquasar jet in NGC 300}

\author[A. F. McLeod et al.]{A. F. McLeod$^{1,2},$\thanks{E-mail: anna.mcleod@berkeley.edu}
S. Scaringi$^{2}$,
R. Soria$^{3,4,5}$,
M. W. Pakull$^{6}$,
R. Urquhart$^{4,7}$,\newauthor
T. J. Maccarone$^{2}$,
C. Knigge$^{8}$,
J. C. A. Miller-Jones$^{4}$,
R. M. Plotkin$^{4}$,
C. Motch$^{6}$,\newauthor
J. M. D. Kruijssen$^{9}$ and A. Schruba$^{10}$
\\
$^{1}$Department of Astronomy, University of California Berkeley, Berkeley, CA 94720, USA\\
$^{2}$Department of Physics \& Astronomy, Texas Tech University, PO Box 41051, Lubbock, TX 79409, USA\\
$^{3}$College of Astronomy and Space Sciences, University of the Chinese Academy of Sciences, Beijing 100049, China\\
$^{4}$International Centre for Radio Astronomy Research, Curtin University, GPO Box U1987, Perth, WA 6845, Australia\\
$^{5}$Sydney Institute for Astronomy, School of Physics A28, The University of Sydney, Sydney, NSW 2006, Australia\\
$^{6}$Observatoire Astronomique, Universit{\'e} de Strasbourg, CNRS, UMR 7550, 11 Rue de l'Universit{\'e}, 67000, Strasbourg, France\\
$^{7}$Center for Data Intensive and Time Domain Astronomy, Department of Physics and Astronomy, Michigan State University, East Lansing, MI, USA\\
$^{8}$School of Physics \& Astronomy, University of Southampton, Southampton SO17 1BJ, UK\\
$^{9}$Astronomisches Rechen-Institut, Zentrum f{\"u}r Astronomie der Universit{\"a}t Heidelberg, M{\"o}nchhofstrasse 12-14, 69120 Heidelberg, Germany\\
$^{10}$Max-Planck-Institut f{\"u}r extraterrestrische Physik, Giessenbachstrasse 1, D-85748 Garching
}

\date{Accepted XXX. Received YYY; in original form ZZZ}

\pubyear{2015}

\hypersetup{draft}
\begin{document}
\label{firstpage}
\pagerange{\pageref{firstpage}--\pageref{lastpage}}
\maketitle

\begin{abstract} 
We recently reported the discovery of a candidate jet-driving microquasar (S10) in the nearby spiral galaxy NGC 300. However, in the absence of kinematic information, we could not reliably determine the jet power or the dynamical age of the jet cavity. Here, we present optical MUSE integral field unit (IFU) observations of S10, which reveal a bipolar line-emitting jet structure surrounding a continuum-emitting central source. The optical jet lobes of S10 have a total extent of $\sim$ 40 pc and a shock velocity of $\sim$ 150 km s$^{-1}$. Together with the jet kinematics, we exploit the MUSE coverage of the Balmer H$\beta$ line to estimate the density of the surrounding matter and therefore compute the jet power to be $P_{jet}\approx$ 6.3 $\times$ 10$^{38}$ erg s$^{-1}$. An optical analysis of a microquasar jet bubble and a consequent robust derivation of the jet power have been possible only in a handful of similar sources. This study therefore adds valuable insight into microquasar jets, and demonstrates the power of optical integral field spectroscopy in identifying and analysing these objects.   
\end{abstract}




\section{Introduction}

Accretion-powered objects of all types appear to produce jets of some kind (see e.g. \citealt{livio1999}). On galactic scales, radio jets are seen in many active galaxies and quasars. On stellar-mass scales, the presence of jets is well established in systems ranging from young stellar objects to accreting neutron stars and black holes. 
In recent years, jets have even been discovered in cataclysmic variables (accreting white dwarf systems, \citealt{koerding}) and accreting X-ray pulsars \citep{vandeneijnden}. Both of these classes had previously been thought to be incapable of driving such powerful collimated outflows. All of this suggests that jets are a key ingredient in the physics of disk accretion. In addition, jets are also important sources of feedback into the interstellar medium, both in terms of total kinetic power, and as sources for the production of cosmic rays (\citealt{heinz02,fender16,romero17}). The term {\it microquasar} \citep{mirabel92} is often used to describe the most extreme X-ray binaries with strong, resolved jets.Two of the best-known Galactic examples which show the most prominent resolved jets are the accreting black holes SS~433 (e.g.~\citealt{zealey80,fabrika04,brinkmann07,goodall11,farnes17}) and GRS~1915+105 (e.g.~\citealt{castro92,mirabel94,reid14,tetarenko18}). Both of these systems are thought to be accreting at rates in excess of the Eddington limit for a $\simeq 10~M_{\odot}$ black hole. 

In external galaxies, searches for stellar-mass black holes accreting at similarly high rates have focused on {\it ultra-luminous X-ray sources} (ULXs: $L_{X} \gtrsim 10^{39}$~erg~s$^{-1}$). At least some ULXs are now known to contain neutron stars (based on the detection of coherent pulsations, e.g.~\citealt{doroshenko15}), implying accretion at even more highly super-Eddington rates and/or strong beaming. Regardless of the nature of the compact accretor, ULXs represent a promising population in which to search for extragalactic microquasars.

Many ULXs in the Local Universe are surrounded by large optical nebulae (e.g. \citealt{pakull02}). In some cases, such as the nebula around Holmberg II X-1 \citep{pakull02}, the observed line ratios suggest that the gas is X-ray photoionised; in other cases, such as the one around NGC\,1313 X-2, collisional ionisation dominates \citep{pakull08}. These nebulae provide important information about the nature and the true energy output of ULXs. In photo-ionised nebulae, the total power emitted in the HeII $\lambda 4686$ line leads
to an estimate of the X-ray luminosity of the central source \citep{pakull02}, which argues strongly against high beaming
factors. In shock-ionised nebulae, also known as ``ULX bubbles'', the power emitted in diagnostic lines, such as H$\beta$, [OIII] $\lambda 5007$, and [FeII] $\lambda 1.64 \mu$m, is a proxy for the input mechanical power \citep{dopita96} and shows that the kinetic output (in the form of jets and/or winds, e.g. \citealt{siwek17}) can be of the same order of magnitude as the radiative output. 

X-ray photoionised nebulae (powered by accreting compact objects, see also \citealt{kallman82,pakull86}) can be easily distinguished from ordinary HII regions (ionised by stellar photons), because of the co-existence of lower-ionisation and higher-ionisation lines of the same elements ({\it e.g.}, [OIII], [OII] and [OI]), and the higher relative strength of high-ionisation lines such as those from HeII and [NeV] \citep{urquhart18}. Shock-ionised ULX bubbles can be distinguished from supernova remnants (SNRs) because of their larger size (diameters of $\approx$ 100--300 pc, e.g. \citealt{pakull08}) and dynamical age ($\sim$ 10$^5$ yr).

Both kind of ULX nebulae (photo- and shock-ionised) provide an additional, independent line of evidence that geometric and relativistic beaming are unlikely to be of strong importance for producing the high luminosities seen in most ULXs \citep{pakull02,kaaret04,binder18}. If high beaming factors were present, for every nebula containing a ULX, there should be a large number of isotropically emitting nebulae without a strong X-ray source in the center (representing systems in which the X-rays are beamed away from our line of sight). This is in contradiction with observational results.
The few nebulae without ultraluminous X-ray sources are then of great interest \citep{pakull10}.  It is likely that these are ULXs viewed edge-on, like SS 433 in the Milky Way, so that the inner accretion flow where most of the X-rays are produced is obscured by the puffed-up outer accretion disk.  Developing good statistics on the number, kinetic power, and eventually the orbital periods of these systems may thus help understand the nature of the ultraluminous sources' geometries.  It is thus of great importance to discover and characterize these nebulae.

In a recent publication, \citealt{urquhart19} (henceforth referred to as Paper I) reported the discovery and carried out a comprehensive multi-wavelength analysis of a candidate microquasar in NGC\,300, associated with the collisionally-ionised nebula S10, which had previously been interpreted as a SNR \citep{dodorico80,blair97}. While not sufficiently bright to classify S10 as a ULX, the {\it Chandra} X-ray data show an elongated ($\sim$ 150 pc) structure consisting of four discrete knots, which is suggested to result from shocks along the axis of a jet as it propagates through the local interstellar medium (ISM). In the optical, {\it Hubble Space Telescope} ({\it HST}) broad-band and Very Large Telescope (VLT) narrow-band images show a bubble and likely jet lobes seemingly associated with the X-ray structure. In the radio regime, 5-GHz and 9-GHz maps from the Australia Telescope Compact Array (ATCA) were also discussed in Paper I, and supported the identification of a large-scale jet, with optically-thin synchrotron emission. In Paper I, we estimate the jet power according to the method of \citet{pakull10}, where the main derived quantity is the shock velocity. From the total H$\beta$ luminosity, we find a jet power P$_{\mathrm{jet}}>10^{39}$ erg s$^{-1}$. However, this result depends on an assumed shock velocity, and a more reliable estimate of the jet power can be obtained by computing the dynamical age of the shocked bubble. For this, the shock velocity, the ambient ISM density and the size of the bubble are required \citep{weaver77,kaiser97,lamers99}.

In this paper, we use the Multi Unit Spectroscopic Explorer (MUSE; \citealt{muse}) on the VLT to analyse the kinematic and physical properties of the S10 nebula. MUSE covers all the main nebular emission lines in the optical wavelength regime ($\approx$ 4750--9350 \AA\ in the nominal mode). The simultaneous spectral and spatial coverage of this integral field spectrograph enables us to determine the kinematics of the shocked gas and the ISM density, and therefore derive a more robust measurement of jet power. Moreover, our MUSE data allow us to determine the spatial orientation of the bipolar jet along the line of sight. This paper is organised as follows. Section \ref{obs} discusses the MUSE observations, data reduction and astrometry; in Section \ref{analysis}, we analyse the morphology, kinematics, nebular emission and stellar counterpart of S10, and compute the jet power; conclusions are presented in Section \ref{conc}.

\begin{figure*}
\mbox{
\subfloat[]{\includegraphics[scale=0.5]{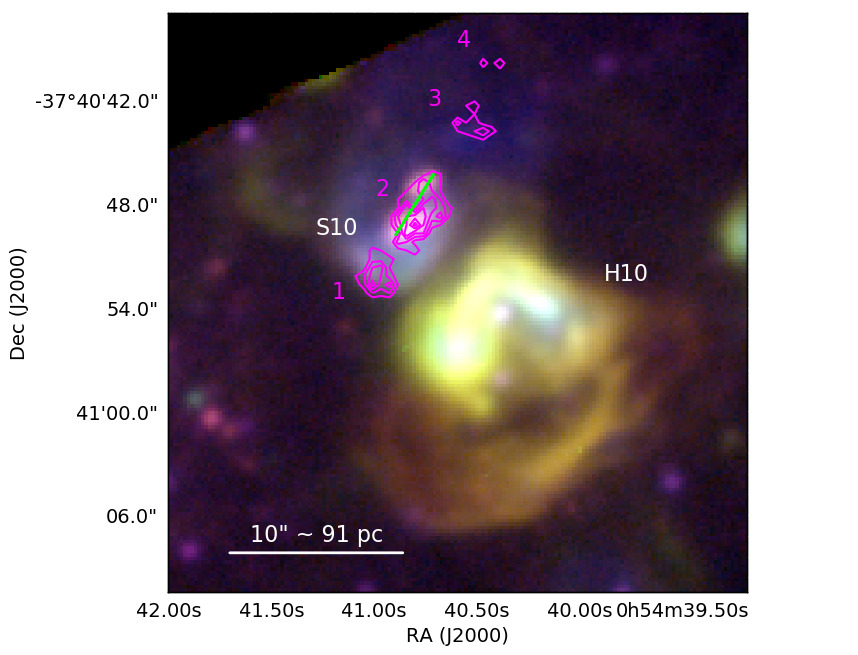}}
\subfloat[]{\includegraphics[scale=0.35,trim=0 -2.3cm 0 0]{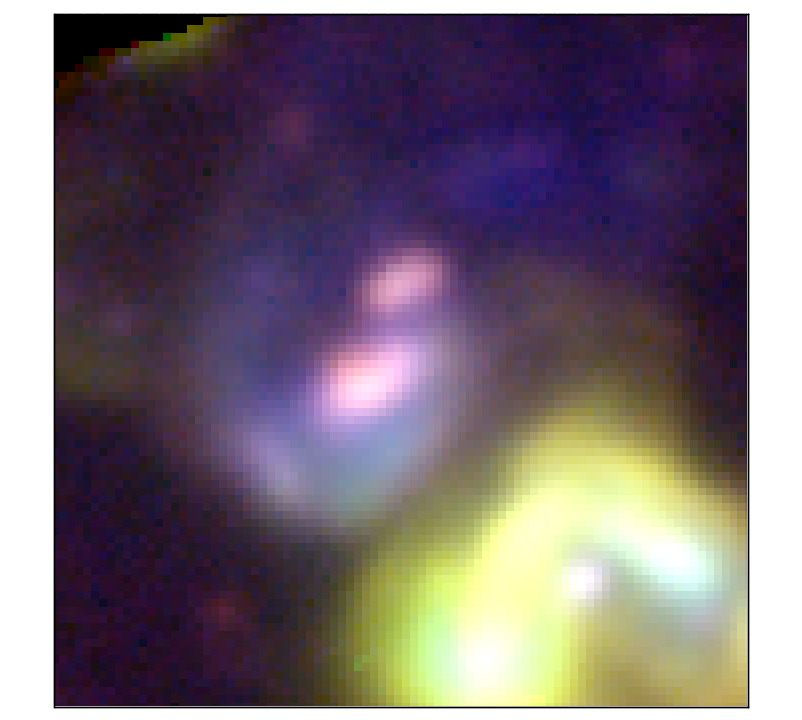}}}
\caption{RGB composite of the S10/H10 region (red = [SII] $\lambda$6717, green = H$\alpha$, blue = [OIII] $\lambda$5007, not background subtracted), with {\it Chandra} X-ray contours tracing the 4 knots overlaid in magenta (see Paper I). The solid green line on X-ray knot 2 indicates the peak-to-peak separation of $\sim$ 40 pc measured from the velocity map (Fig. \ref{s2_vel}). North is up and East is left. The optical nebula of S10 coincides with the largest X-ray knot. For clarity, panel (b) is a 0$'$.3 $\times$ 0$'$.3 zoom-in of S10.}
\label{rgb}
\end{figure*}

\section{Observations}\label{obs}
The data presented in this work were taken with the integral field spectrograph MUSE (mounted on the VLT), as part of the observing program 098.B-0193(A) (PI McLeod). This program consists of a contiguous 35-pointing mosaic covering the NGC 300 disk. Here, we only analyse data from the MUSE cube covering NGC 300-S10. The data cube is a combination of three 900 seconds integration time single telescope pointings taken in a 90$^{\circ}$ rotation dither pattern, a strategy which has proven successful in reducing instrumental artefacts in our previous MUSE programs (e.g. \citealt{m16}, \citealt{pillars}).
The observations were carried out in January 2017 under grade A observing conditions, with an average seeing of 0$''$.5. This program was taken in the Wide Field Mode of MUSE, with a pixel scale of 0.2 arcsec/pixel, a resolving power of 1770 to 3590 (from 4750 to 9350 \AA\ ), and a spectral separation of 1.25 \AA\ between single frames.

The data reduction was carried out with the MUSE pipeline \citep{pipeline} in the {\sc esorex} environment using the standard calibration files for that particular night and the static calibrations of the pipeline. Emission line and ratio maps are obtained as in our previous MUSE programs (\citealt{m16}, \citealt{orion}, \citealt{pillars}), and a MUSE three-color composite of the [SII]$\lambda$6717, H$\alpha$ and [OIII]$\lambda$5007 lines is shown in Fig. \ref{rgb}. The galactic background (selected from an emission-free region) is subtracted from each integrated emission line map, and we mask pixels with negative values in the map with lowest S/N ([NII]$\lambda$6548).

Because of the 90$^{\circ}$ rotation dither pattern used for our observations, we want to exclude the introduction of world coordinate system (WCS) shifts/distortions which would complicate the overlay with the X-ray data in Section \ref{analysis} and the {\it HST} broad-band filter image in Section \ref{stellar}. For this we compare the MUSE WCS to that of {\it HST} data by using the {\sc photutils} package \citep{photutils} to identify point sources in both the archival {\it HST} F814W and the MUSE V band images. Here, the V band filter image was obtained by collapsing the combined MUSE cube over the Johnson-Cousins V band filter specifications, with a central wavelength of 5477 \AA\ and a FWHM of 991 \AA. In a second step, we cross match the two point source catalogs, obtaining a mean separation of $\sim$ 0$''$.12 between cross-matched objects, smaller than the MUSE pixel scale of 0$''$.2. We are therefore confident in adopting the MUSE WCS for the purpose of the analyses presented in this paper. The cross-matched sources are shown in the Appendix, Fig. \ref{star}(a) and \ref{star}(b) for {\it HST} and MUSE, respectively.

\section{Analysis}\label{analysis}

\subsection{Morphology and kinematics}\label{morphkin}
In Paper I, the presence of a powerful microquasar is inferred from a string of four aligned X-ray knots (magenta contours in Fig. \ref{rgb}) and an elongated radio structure, associated with the shock-ionised optical nebula S10 (RA $= 00^{h} 54^{m} 40^{s}.87$, Dec $= -37^{\circ}40'48''.73$, J2000). As part of one of the earliest classifications of this source, \citet{blair97} identify S10 as a candidate supernova remnant. Immediately South-West of S10 is another large optical nebula, labeled H10 in \cite{blair97}, and interpreted as an ordinary H II region.

Here, we exploit the MUSE coverage of the main optical nebular emission lines to analyse the morphology and kinematics of the ionised gas associated with S10 and H10. Fig. \ref{rgb} shows a three-color composite of the region of interest, where red is [SII]$\lambda$6717, green is H$\alpha$, and blue corresponds to [OIII]$\lambda$5007. S10 is clearly identified in the MUSE data, with two bright emission line regions coincident with X-ray knot 2 (magenta contours). This indicates that X-ray knot 2 is indeed extended, as suggested in Paper I. The MUSE data also reveal a bow-shock-like structure to the South-East (part of which is coincident with knot 1), as well as diffuse [OIII] emission (blue in Fig. \ref{rgb}) towards the northern end of S10. The HII region H10 is a bubble-shaped structure, traced by bright emission on its northern rim (facing S10), and a more filamentary southern rim. Integrated flux values of the main nebular emission lines for S10 and H10 are listed in Table \ref{fluxes}. 

\begin{figure*}
\mbox{
\subfloat[]{\includegraphics[scale=0.45]{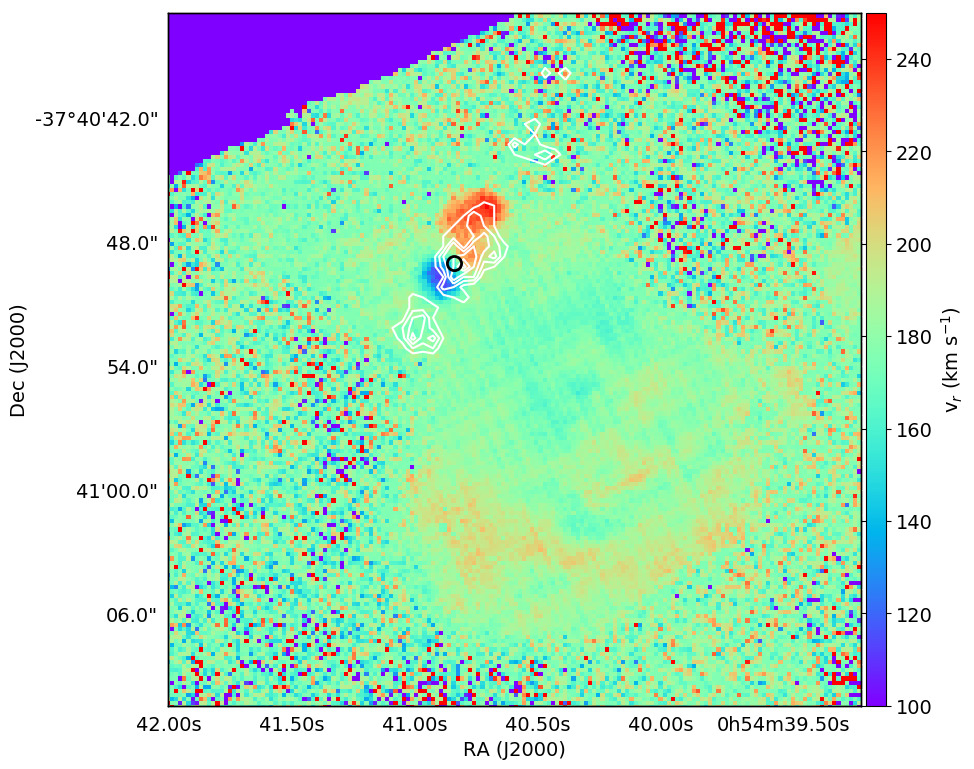}}
\subfloat[]{\includegraphics[scale=0.35,trim=0 -2.3cm 0 0]{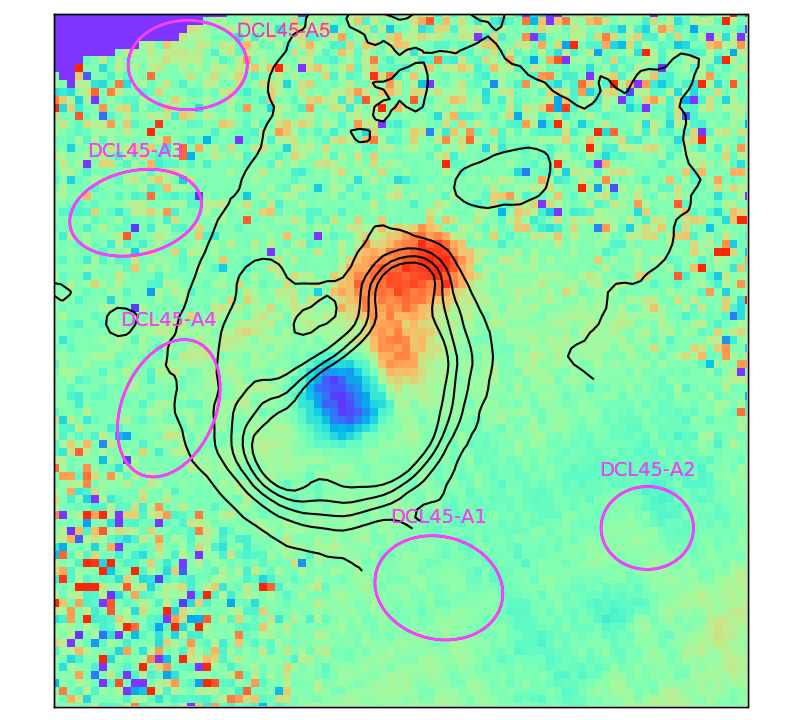}}}
\caption{Panel (a): [SII]$\lambda$6717 radial velocity map with {\it Chandra} X-ray contours in white, the black circle shows the location of star C (see Paper I and Section \ref{stellar}) Panel (b): same as panel (a), zoomed in 0$'$.3 $\times$ 0$'$.3 around S10. Black contours map the observed [OIII]$\lambda$5007 flux (ranging from 1.96$\times$10$^{-17}$ to 3.2$\times$10$^{-17}$ erg cm$^{-1}$ s$^{-2}$ pixel$^{-1}$), which traces the ionized gas in the bow-shock (Fig.~\ref{rgb}); the magenta regions indicate CO intensity peaks (from the ALMA images of \citealt{faesi18}).}
\label{s2_vel}
\end{figure*}

\begin{figure}
\hspace{-1cm}
\includegraphics[scale=0.6]{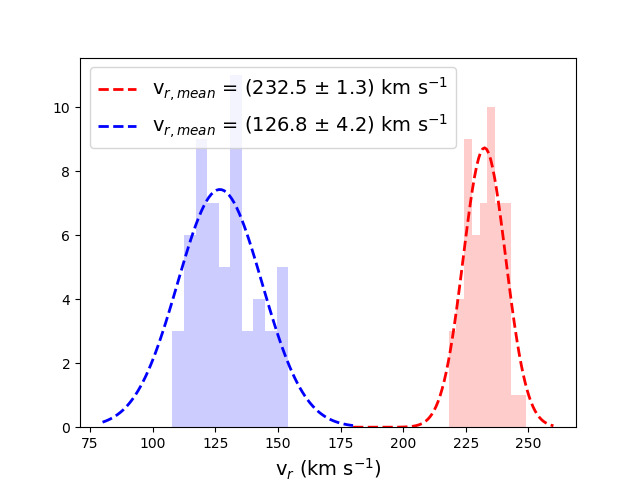}
\caption{Normalized histograms of radial velocity pixel values fro the [SII]$\lambda$6717 rest frame (see figures \ref{s2_vel} and \ref{jet_mask}) representative of the blue and red jet lobes, and Gaussian fits to the distributions.}
\label{vel_hist}
\end{figure}

\begin{figure}
\mbox{
\subfloat[]{\includegraphics[scale=0.33]{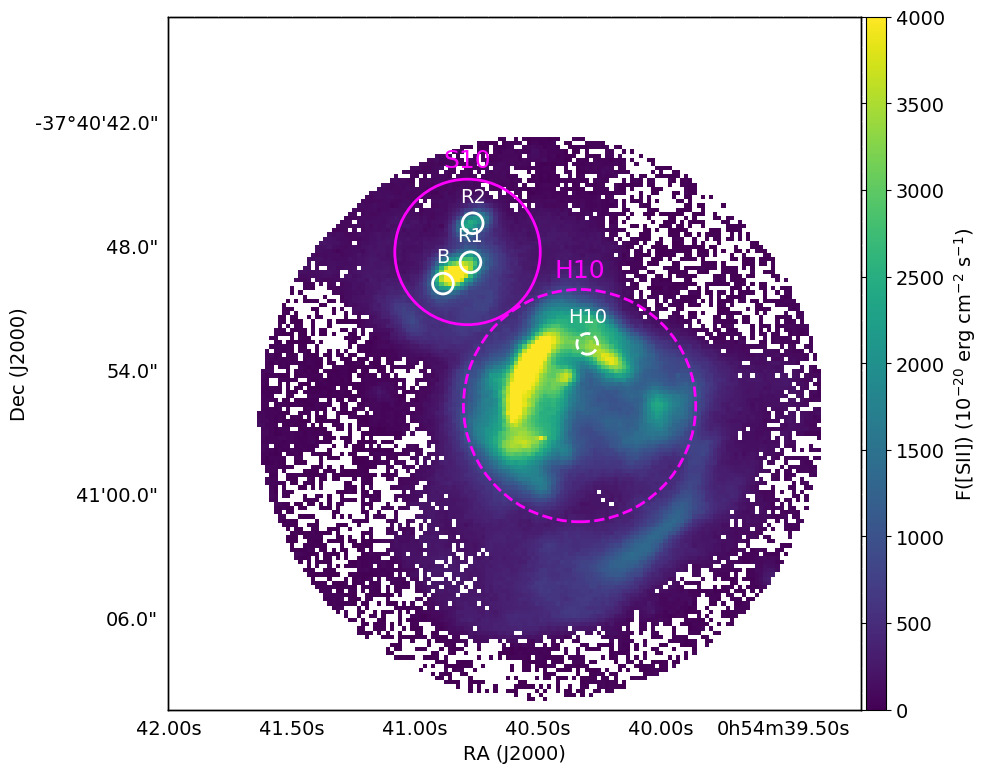}}}
\mbox{
\subfloat[]{\includegraphics[scale=0.5]{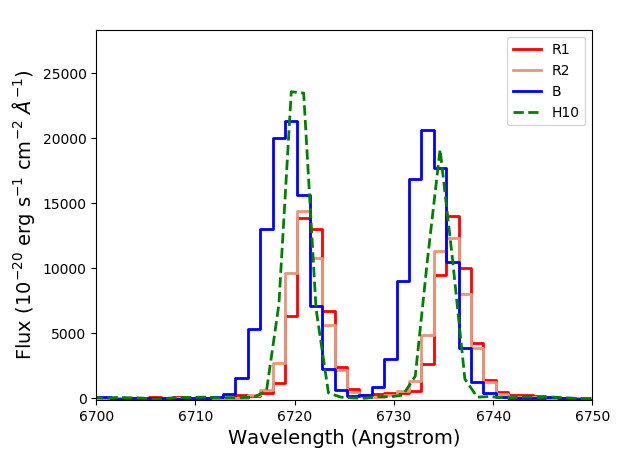}}}
\mbox{
\subfloat[]{\includegraphics[scale=0.5]{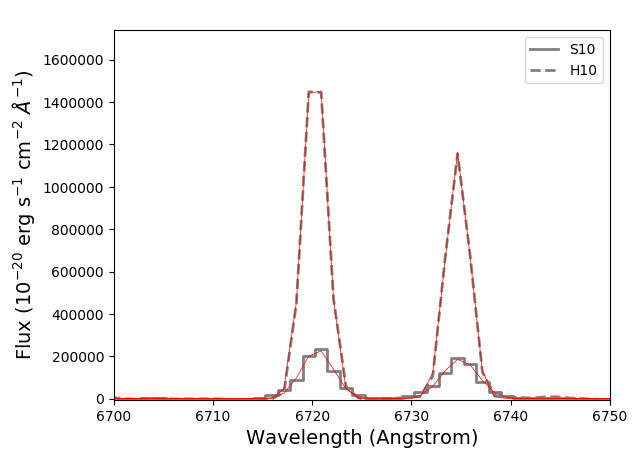}}}
\caption{Panel (a): [SII]$\lambda$6717 integrated line map. The white circles indicate the regions used to extract the spectra shown in panel (b), while the magenta circles show the regions used to extract integrated spectra for S10 and H10, shown in panel (c). Dashed circles refer to H10, solid circles to S10. Panel (b): continuum-subtracted spectra of the blue- and redshifted lobes and H10 cropped around the [SII]$\lambda$6717,31 lines, extracted from the white regions shown in panel (a). Panel (c): same as panel (b) for the magenta regions, red lines are Gaussian fits (see Table \ref{specfit}).}
\label{s2_spec}
\end{figure}

\begin{table}
\begin{center}
\caption{Emission line fluxes relative to H$\alpha$ (set to 300 units) extracted from the magenta circular regions around S10 and H10 shown in Fig. \ref{s2_spec}(a).}
\begin{tabular}{lcc}
\hline 
\hline
Line & \multicolumn{2}{c}{Flux}\\
\hline
& S10 & H10 \\
\hline
H$\beta$ & 98 & 95\\
5007 [OIII] & 206 & 38 \\
6548 [NII] & 38 & 24\\
H$\alpha$ & 300 & 300\\
6584 [NII] & 115 & 68\\
6717 [SII] & 116 & 62\\
6731 [SII] & 107 & 45\\
\hline
\hline
\label{fluxes}
\end{tabular}
\end{center}
\end{table}

\begin{table*}
\begin{center}
\caption{Peak fluxes (amplitudes) and widths from Gaussian fits (line+continuum) to the integrated spectra shown in Fig. \ref{s2_spec}(c), extracted from the magenta circles in panel (a) of the same figure. Errors on the amplitudes are of the order of $\sim$ 3--4$\times$10$^{-20}$ erg s$^{-1}$ cm$^{-2}$ \AA$^{-1}$, and errors on the widths are of the order of $\sim$ 0.03 \AA.}
\begin{tabular}{lccccc}
\hline 
\hline
Region & A$_{6717}$ & $\sigma_{6717}$ & A$_{6731}$ & $\sigma_{6731}$ &  N$_{\rm{e}}$\\
 & (10$^{-16}$ erg s$^{-1}$ cm$^{-2}$ \AA$^{-1}$) & \AA & (10$^{-16}$ erg s$^{-1}$ cm$^{-2}$ \AA$^{-1}$) & \AA & (cm$^{-3}$) \\
\hline
S10 & 23.2 & 1.60 & 18.7 & 1.80 & 456 \\
H10 & 167.0 & 1.18 & 116.2 & 1.17 & 42\\
\hline
\hline
\label{specfit}
\end{tabular}
\end{center}
\end{table*}

The kinematics of the ionised gas are illustrated in Fig. \ref{s2_vel}, which shows a radial velocity map obtained by applying a pixel-by-pixel single component Gaussian fitting routine to the [SII]$\lambda$6717 line. The southern H10 shell can be identified at velocities of $\approx$ 195 km s$^{-1}$, in good agreement with the radial velocity map of NGC 300 obtained by \cite{westmeier11} from HI 21 cm data. The striking feature in Fig. \ref{s2_vel}, however, is coincident with X-ray knot 2 and consists of a bipolar structure aligned with the general direction of S10. Together with the results of Paper I, this bipolar morphology of S10 confirms the presence of a microquasar jet powered by accretion onto a compact central object and traced by the emission of the ionised bubble driven by the bipolar jet. Given the distance to NGC 300 of $\sim$ 2 Mpc \citep{dalcanton09}, we infer the ionised shocks to be $\approx 40.1$ pc apart (i.e. the projected separation between the two velocity peaks, see also Fig.~\ref{rgb}). We use {\sc glue}\footnote{http://glueviz.org/} to extract representative pixel values of the red and blue lobes of S10 (see Appendix, Fig. \ref{jet_mask}), allowing us to further analyse the bipolar jet bubble. Histograms of the extracted pixel velocities are shown in Fig. \ref{vel_hist}. Gaussian fits to the distributions yield a projected differential velocity $\Delta v \sim$ (105 $\pm$ 4) km s$^{-1}$ between the red-shifted and blue-shifted peaks. Thus, the forward-shock velocity is $>$ 50 km s$^{-1}$, which however is a strict lower limit given that projection effects are playing a significant role and the absolute shock velocity is potentially much higher (see Section \ref{ratios}). To further illustrate the shifted line centroids of the two lobes we plot their [SII] spectra in Fig. \ref{s2_spec}(b) (where we label the blue lobe as B, and two regions in the red lobe as R1 and R2), together with a spectrum of H10.

The velocity structure of S10 shows a clear asymmetry between the red and blue lobes where the red lobe is almost double in angular size than the blue lobe (see Fig.~\ref{s2_vel}). This is a possible indication for a scenario in which the blue lobe is moving towards (and therefore being slowed down by) a region of higher density as compared to the redshifted region. If so, we would expect to find molecular material in the path of the blue lobe. To test this scenario we retrieve the coordinates, position angles and approximate sizes of ALMA CO intensity peaks in the region from \cite{faesi18} and plot them in Fig. \ref{s2_vel}(a). We find that the CO peaks trace the bow-shock shaped structure seen in Fig. \ref{rgb}(b), therefore supporting the picture in which the blueshifted jet lobe is expanding into a region higher density, where it is being slowed down and a bow-shock is observed.

\subsection{Emission line ratios}\label{ratios}
\begin{figure}
\hspace{-0.2cm}
\includegraphics[scale=0.6]{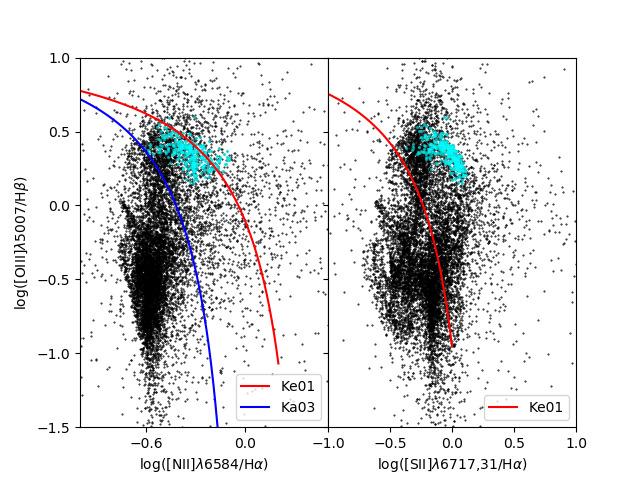}
\caption{BPT diagram of the analysed MUSE cube, each data point corresponds to a pixel. The red line corresponds to the \citet{kewley01} line dividing AGN from HII regions, while the blue line corresponds to the \citet{kauffmann03} line. The latter accounts for galaxies whose emission lines contain contributions from both star formation and AGN. The cyan data points correspond to S10.}
\label{bpt_fig}
\end{figure}

We exploit MUSE's coverage of the optical nebular emission lines to further investigate the physical conditions of the ionised gas in terms of shock- vs. photo-ionisation and ISM densities. In Paper I, we suggested that the emission in S10 is mainly due to shock-ionisation, as opposed to photo-ionisation (which is responsible for part of the emission from H10). Here, we further strengthen this statement with the diagnostic BPT diagram \citep*{bpt}, which compares ratios of the main collisionally excited lines to the Balmer recombination lines. Shock- and photo-ionisation-dominated regions occupy different parts of the BPT diagram, and it has therefore been used extensively to distinguish between HII regions and AGN, for example. Generally, one distinguishes between HII regions (photo-ionisation) and AGN-dominated galaxies (shock-ionisation) depending on whether a source is found above or below the {\it maximum starburst line}, as determined from photoionisation and stellar synthesis population models \citep{kewley01}. The [NII] and [SII] BPT diagrams for the FOV covered by the data cube are shown in Fig. \ref{bpt_fig}, where the data points corresponding to S10 are marked in cyan. The S10 data points lie above the maximum starburst lines in both the [SII] \citep{kewley01} and the [NII] \citep{kauffmann03} BPT diagrams (where the blue line in the [NII] BTP diagram corresponds to a modified \citealt{kewley01} line which also accounts for composite HII region and AGN excitation), suggesting that shocks (or non-stellar AGN-type ionising continua) are the dominant ionisation mechanism. The BPT diagram line ratios of the individual pixels are indicative of shock velocities of $v_{\rm s}\sim$ 55 -- 90 km s$^{-1}$, which correspond to a lower limit. Below we derive shock velocities of the red- and blueshifted lobes from the FWHM of their integrated spectra.

Further line ratios are shown in Figures \ref{line_ratios}, where the ratio of the [SII] lines traces the electron density and enhanced values of the [SII]$\lambda$6717/H$\beta$ ratio trace shock-ionized gas. It should be acknowledged that -- as also mentioned in \citealt{m16} and \citealt{ercolano12} -- some caution is required when applying BPT diagnostics to the emission line properties of spatially resolved regions. Specifically, the models used to interpret the BPT diagram predict integrated fluxes for single galaxies or HII regions, whereas our spatially resolved emission line ratios will reflect local variations of physical properties of the gas such as the density and the temperature \citep{orion}. A more detailed emission line ratio and shock analysis of S10 and other shock-ionised sources in NGC 300 will be discussed in a forthcoming publication.

The derivation of the jet power (see Section \ref{power}) relies on the knowledge of the density of the pre-shock (ambient) ISM into which the jet is propagating. The ISM density does not correspond to the density as one would measure e.g.~from the ratio of the [SII] lines, as this would deliver the electron density of the shocked matter. Moreover, local variations of physical parameters, together with weak emission line intensities, can lead to unrealistically high (or low) values when deriving e.g. the electron temperature on spatially-resolved scales. To estimate the pre-shock density $n_{\rm{ISM}}$, we therefore use the surface brightness of the H$\beta$ line according to \cite{dopita96}, who give a relation between the flux of the H$\beta$ line from a surface element of a radiative shock with velocity $v_{\rm{s}}$ and the density,

\begin{equation}\label{dopita}
    f_{\rm{H\beta}} = 7.44\times10^{-6}v_{2}^{2.41}\Bigg(\frac{n_{\rm{ISM}}}{\rm{cm^{-3}}}\Bigg)~~\rm{erg~s^{-1}~cm^{-2}}
\end{equation}

\noindent where $v_{2}$ corresponds to $v_{\rm{s}}$ in units of 100 km s$^{-1}$. A bubble of radius R at a distance D from the observer spans a solid angle of $\Omega=\pi (R/D)^{2}$ on the sky and is observed with a flux $F_{\rm{H\beta}} = L/(4 \pi D^{2}) = 4 \pi R^{2} f_{\rm{H\beta}} / (4 \pi D^{2})$ and an intensity $I_{\rm{H\beta}} = F_{\rm{H\beta}}/\Omega = f_{\rm{H\beta}}/\pi$. Hence, Eq. \ref{dopita} becomes

\begin{equation}
    I_{\rm{H\beta}} = 2.37\times10^{-6}v_{2}^{2.41}\Bigg(\frac{n_{\rm{ISM}}}{\rm{cm^{-3}}}\Bigg)~~\rm{erg~s^{-1}~cm^{-2}~sr^{-1}}
\end{equation}

\begin{figure}
\includegraphics[scale=0.35]{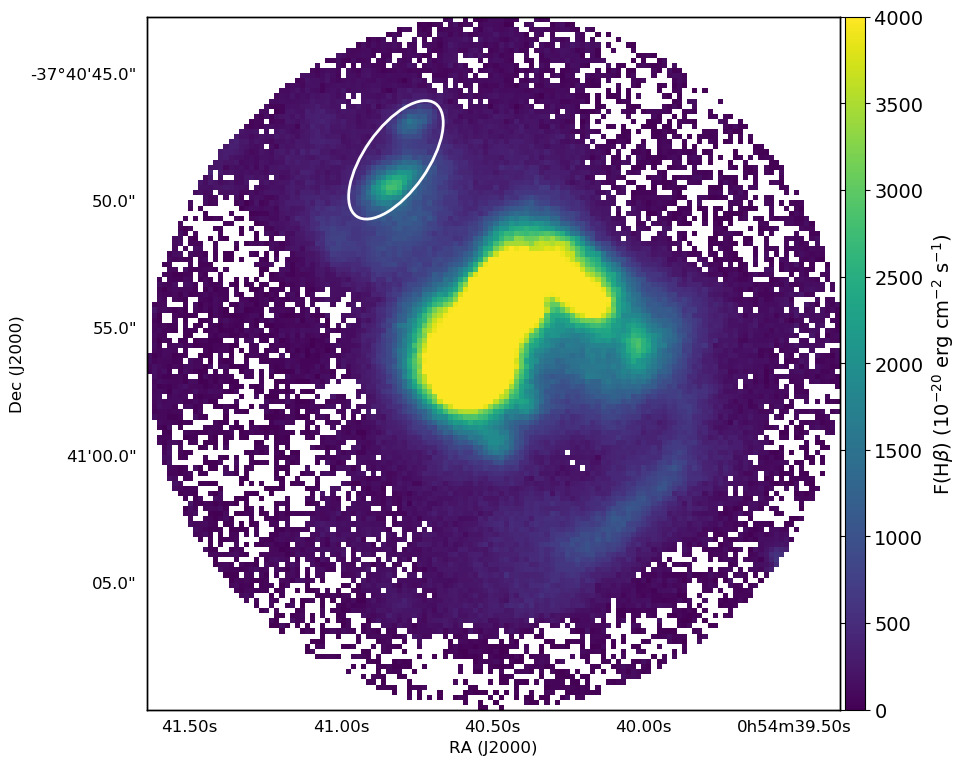}
\caption{The H$\beta$ integrated line map, the white ellipse shows the region used to derive the H$\beta$ flux, see Section \ref{ratios}.}
\label{hbeta}
\end{figure}

For the spectra shown in Fig.~\ref{s2_spec}(b) we measure FWHM of 4.1, 4.2, 5.1 and 3.3 \AA\ for R1, R2, B and H10, respectively. With an instrumental FWHM of $\approx$ 3 \AA\ of MUSE at 6700 \AA\, the mean intrinsic line width of the S10 [SII] lines therefore is $\sim$ 4.5 \AA. This corresponds to shock velocities $v_{\rm{s}}\sim$ 150 km s$^{-1}$ (with $v_{\rm{s}}\approx$ FWHM, \citealt{heng10}), hence $v_{2} = 1.5$. From the H$\beta$ integrated line map (see Fig.~\ref{hbeta}) we derive $I_{\rm{H\beta}}\simeq$ 1.2$\times 10^{-5}$ erg cm$^{-2}$ s$^{-1}$ sr$^{-1}$, and we therefore obtain $n_{\rm{ISM}}\sim$ 2.0 cm$^{-3}$.

\subsection{The central binary system}\label{stellar}
\begin{figure}
\includegraphics[scale=0.35]{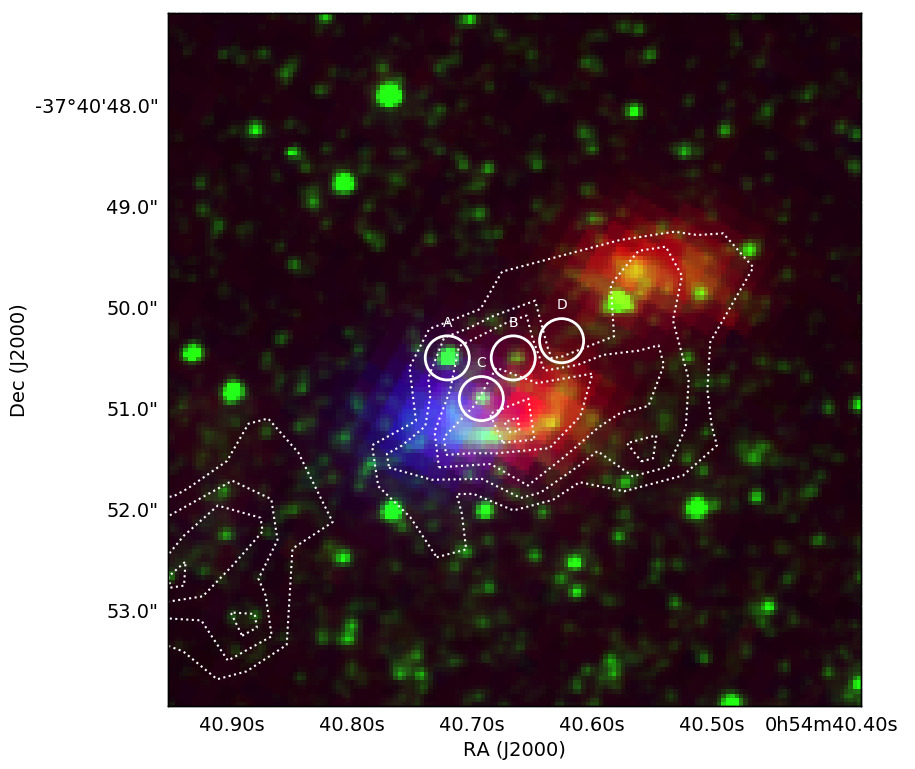}
\caption{RGB composite of the {\it HST} F606W image (green) and two 1 \AA\ wide integrated maps red and blue of the central [SII]6717 line, centered on 6723.4 and 6717.2 \AA, respectively. White circles correspond to the 4 potential stellar counterparts discussed in Paper I, while the white dotted contours trace the X-ray emission.}
\label{star_counter}
\end{figure}

In Paper I, we discuss 4 potential point-like optical counterparts of the compact object that powers S10, selected from {\it HST}/ACS-WFC images (specifically, broadband F814W, F606W and F475W images), and suggest sources C and D (see Fig.~\ref{star_counter}) as the most likely candidates out of the sample of four.

Here, we spatially associate the two jet lobes and the four potential counterparts identified in Paper I. This is shown in Fig. \ref{star_counter}, which consists of a 3-colour composite of the archival {\it HST} F606W image (in green), and two 1 \AA\ wide integrated maps red and blue of the central [SII]$\lambda$6717 line extracted from the MUSE cube, centered on 6723.4 and 6717.2 \AA, respectively. Based on the (projected) positions of the four sources with respect to the bipolar jet shown Fig. \ref{star_counter} (we note that source D is very faint in the F606W image, being a very red source as can be seen in the F814W image shown in Paper I), we suggest that source C is the most likely optical counterpart of  the central binary associated with the S10 microquasar, as it lies almost precisely in between the red and blue jet lobes. By comparing the photometric properties of source C to the Padova isochrones (\citealt{bressan12}, see Paper I), we find this source to likely be an intermediate-age AGB star of $\sim$ 3 M$_{\odot}$. Unfortunately the MUSE data are not deep enough to allow a spectral analysis of this source, but it is detected as a continuum source in the image obtained by collapsing the entire cube along the wavelength axis (not shown here). If source C is indeed the companion of stellar-mass black hole, and the isochrone adequately models the source (i.e. the source's photometric properties are not too strongly affected by accretion disk light), this could indicate Case C mass transfer (i.e. mass transfer which does not begin until after the ignition of helium).  Alternatively, the source could also be wind-fed.

\subsection{Jet Power}\label{power}
\cite{kaiser97} have laid out an analytical model describing the properties of extragalactic radio sources. This model can be applied to microquasars, and more generally to accreting stellar-mass black holes producing jets (e.g. \citealt{gallo05}). By balancing the interior pressure exerted by the shock-ionised jet lobes and the ram pressure of the shocked ISM, it has been shown (\citealt{kaiser97}) that the jet length within the jet lobes grows with time $t$ in such a way that 

\begin{equation}
t \approx L^{\frac{5}{3}} \left(\frac{\rho_0}{P_{\rm jet}}\right)^{\frac{1}{3}}
\end{equation}

\vspace{0.3cm}
\noindent where $L$ is the separation between the accreting source and the jet shocks, and $\rho_0$ is the mass density of the surrounding un-shocked gas. Since the speed of the shocked bubble (i.e. the jet lobe), 

\begin{equation}
v_{bubble} \approx \frac{3}{5}\left(\frac{L^2\rho_0}{P_{\rm jet}}\right)^{-\frac{1}{3}}
\end{equation}

\vspace{0.3cm}
\noindent it follows that the jet lifetime can be estimated as 

\begin{equation}
t = \left(\frac{3}{5}\right)\left(\frac{L}{v_{\rm bubble}}\right)
\end{equation}

\vspace{0.3cm}

The MUSE data reveal the jet size and bubble velocity to be $\approx 40$ pc (from peak to peak) and $\approx 150$ km/s, respectively. This translates to a jet lifetime of $t\approx0.2$ Myr. We can now use the inferred ISM density $n_{\rm{ISM}}\sim 2$ cm$^{-3}$ to estimate a mass density of $\rho_0 \simeq 5.3 \times 10^{-24}$ g cm$^{-3}$. The resulting time-averaged jet power for S10 is then $P_{\rm jet} \approx 6.3 \times 10^{38}$ erg s$^{-1}$, with a total deposited energy in the jet of $E_{\rm jet} \approx 1.6 \times 10^{51}$~erg. For these estimates we have taken into account an extra factor of 2 from the equations of \cite{kaiser97} taking into account both jet lobes observed in S10. Furthermore, we have assumed a relatively wide jet opening angle for all of the above estimates by setting the $c_1$ constant appearing in Equation 4 of \cite{kaiser97} to 1. Although this is not known to be the case for S10, this assumption results in the simplest case. Calculating different $c_1$ values for different scenarios (e.g. spherical lobe) would yield a factor $\approx$ 4 difference in the jet power estimates, well within observational uncertainties such as measured lobe velocities or binary inclination, and in agreement with the jet power estimate reported in Paper I.

Although S10 possesses a smaller jet, lower shock velocities (likely due to S10's propagation into a higher ambient density), and higher surrounding electron density compared to other super-Eddington microquasars (such as NGC 5408 X-1; \citealt{soria06}, IC 342 X-1 and Ho II X-1; \citealt{cseh12}), its total jet power is comparable. Furthermore, we note that the derived timescale of 0.2 Myr is consistent with the origin of the X-ray emission discussed in Paper I, where we argue that the emission is more consistent with a thermal plasma model with a thermal cooling timescale of $\sim$ 570,000 yr rather than synchrotron, as the latter would imply a cooling timescale of $\sim$ 3000 yr.

\section{Conclusions}\label{conc}

We have obtained spatially resolved spectroscopy with VLT/MUSE of the X-ray source and candidate micro-quasar S10 in the nearby galaxy NGC~300. The main results of our analysis of this data are as follows.
\begin{trivlist}
\item[(i)] The bipolar radio and X-ray emission surrounding the X-ray source is associated with a spatially resolved optical emission line region in our data.
\item[(ii)] We kinematically confirm this bipolar structure as a jet from S10, via the detection of distinct blue- and red-shifted ([SII]) emission line signatures with a shock velocity (as derived from the FWHM of the [SII] emission lines) $v_{\rm s} \sim 150$ km s$^{-1}$.
\item[(iii)] The spatial extent of the jet is $\simeq 40$~pc ($L \simeq 20$~pc), and the pre-shock (ISM) density in the material surrounding the jet is $n_{\rm{ISM}}\sim$ 2 cm$^{-3}$ (as measured from the H$\beta$ surface brightness).
\item[(iv)] Combining these estimates, we infer jet lifetime of $t \simeq 0.2~$Myr, a total kinetic jet power of $P_{\rm jet} \approx 6.3 \times 10^{38}$~erg~s$^{-1}$, and a total energy associated with the jet bubble of $E_{\rm jet} \approx 1.6 \times 10^{51}$~erg. 
\item[(v)] Based on these numbers, the jet from S10 is relatively small and slow relative to those in other extragalactic micro-quasars, but its total kinetic power is comparable. The jet is being driven into higher-density matter, which results in lower velocities.
\item[(vi)] We identify the likely optical counterpart of the central binary system as a continuum-emitting source located centrally between the two jet lobes. The SED of the counterpart matches that of a $\simeq 3~M_{\odot}$ AGB star (based on {\it HST} multi-band photometry), suggesting Case~C mass transfer as the mechanism for driving the high accretion rate onto the compact object in S10.
\end{trivlist}

\section*{Acknowledgements}
This research is partly supported by a Marsden Grant from the Royal Society of New Zealand (AFM), and it is based on observations made with ESO Telescopes at the Paranal Observatory under program ID 098.B-0193. JMDK gratefully acknowledges funding from the German Research Foundation (DFG) in the form of an Emmy Noether Research Group (grant number KR4801/1-1) and from the European Research Council (ERC) under the European Union's Horizon 2020 research and innovation programme via the ERC Starting Grant MUSTANG (grant agreement number 714907). RS thanks the Observatoire de Strasbourg for their hospitality during part of this work. JCAM-J is the recipient of an Australian Research Council Future Fellowship (FT140101082). Furthermore, this research made use of Astropy,\footnote{http://www.astropy.org} a community-developed core Python package for Astronomy \citep{astropy:2013, astropy:2018}, Pyspeckit \citep{pyspeckit} and APLpy, an open-source plotting package for Python \citep{aplpy}.



\nocite{*}
\bibliographystyle{mnras}
\bibliography{ngc300}



\appendix

\section{MUSE World Coordinate System}
\begin{figure*}
\mbox{
\subfloat[]{\includegraphics[scale=0.58]{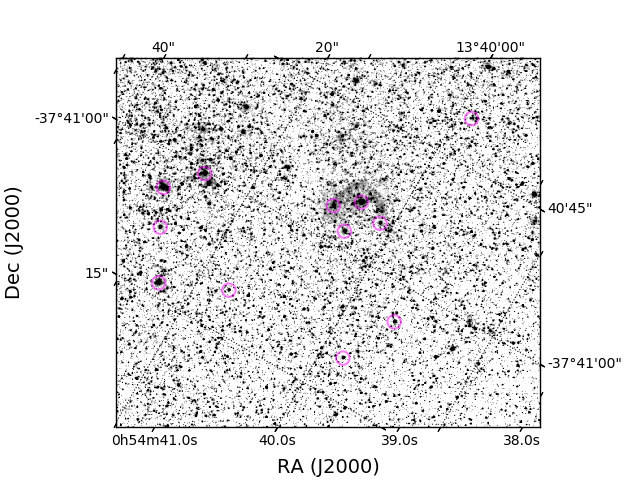}}
\subfloat[]{\includegraphics[scale=0.61]{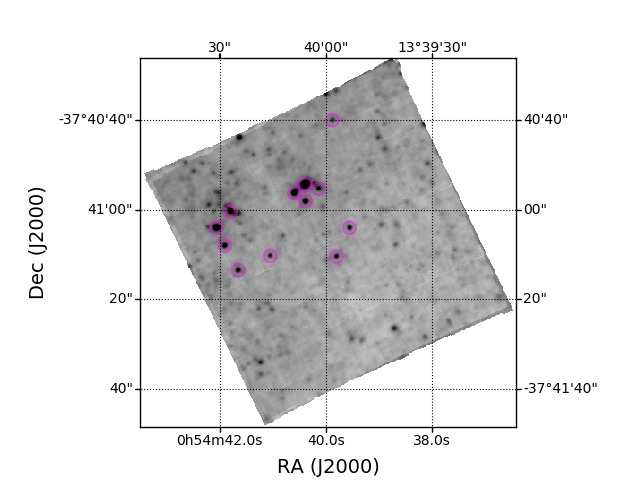}}}
\caption{Left: {\it HST}/F814W image. Right: Johnson-Cousins V band image from the MUSE cube. Magenta circles indicate the position of the cross-matched stars (see Section \ref{obs}).}
\label{star}
\end{figure*}

As described in Section \ref{obs}, we compare the MUSE WCS to that of archival {\it HST} data. This is shown in Fig. \ref{star}, where the cross-matched sources between the two data sets are shown in magenta.

\section{{\sc glue} pixel masks}
\begin{figure*}
\includegraphics[scale=0.4]{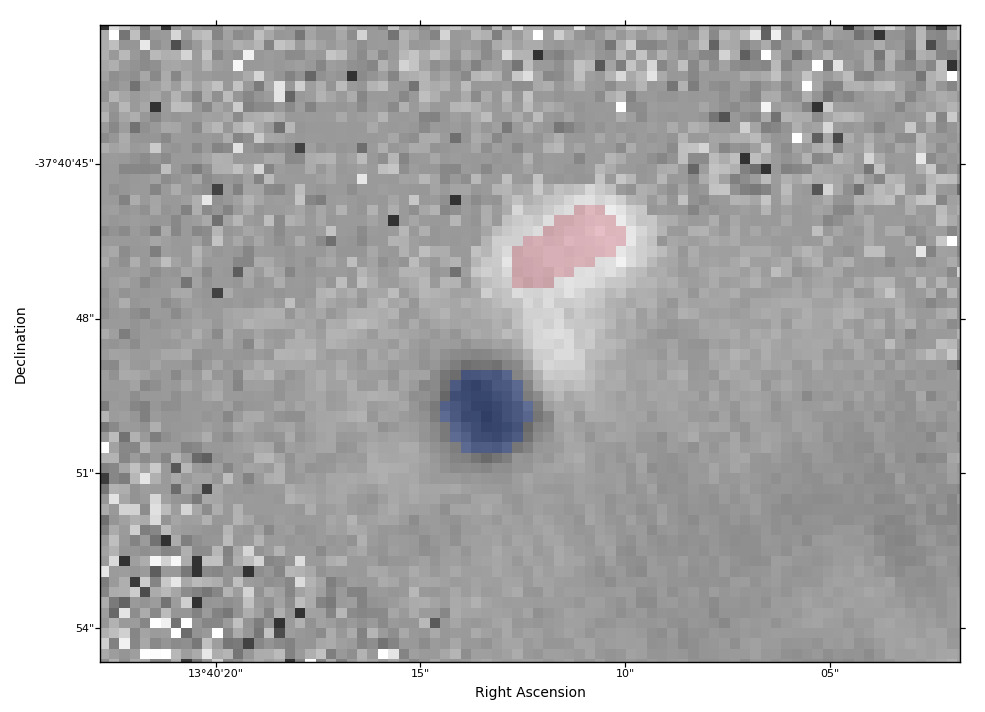}
\caption{Velocity map of the S10 region (scaled to the same minimum and maximum values as in Fig. \ref{s2_vel}) showing the pixels extracted with {\sc glue} to produce Fig. \ref{vel_hist} (see main text Section \ref{analysis}).}
\label{jet_mask}
\end{figure*}

We use {\sc glue} (http://glueviz.org/) to extract pixel values from the velocity of the S10 jet lobes to evaluate their relative velocities, as is discussed in Section \ref{analysis}. This is shown in the [SII] velocity map in Fig. \ref{jet_mask}, which indicates the red/blue regions used for the pixel extraction. Pixel values withing the marked regions are then used as representative values to produce Fig. \ref{vel_hist}.  

\section{Emission line ratios}\label{ratios_appendix}
\begin{figure*}
\mbox{
\subfloat[]{\includegraphics[scale=0.42]{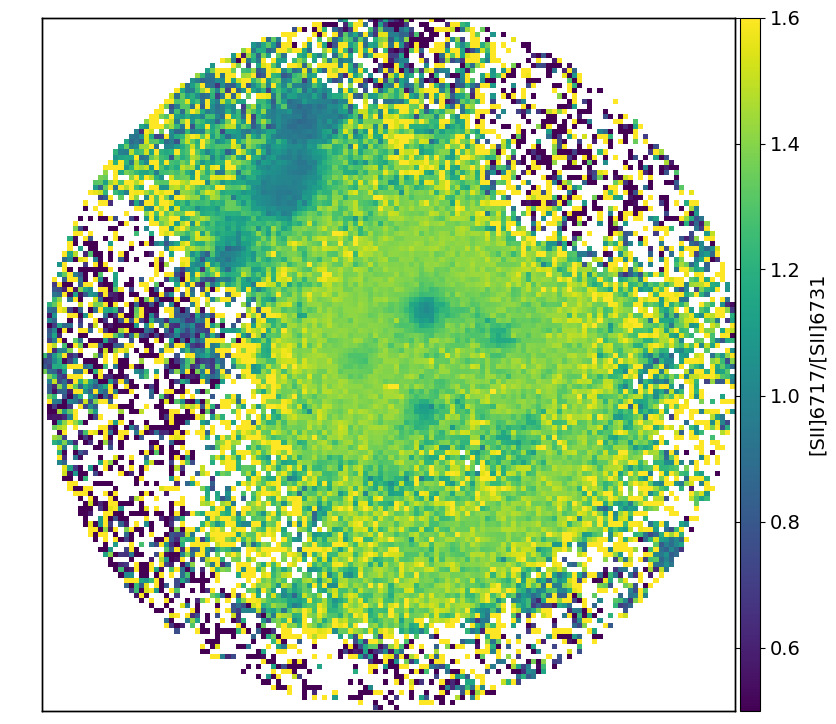}}
\subfloat[]{\includegraphics[scale=0.42]{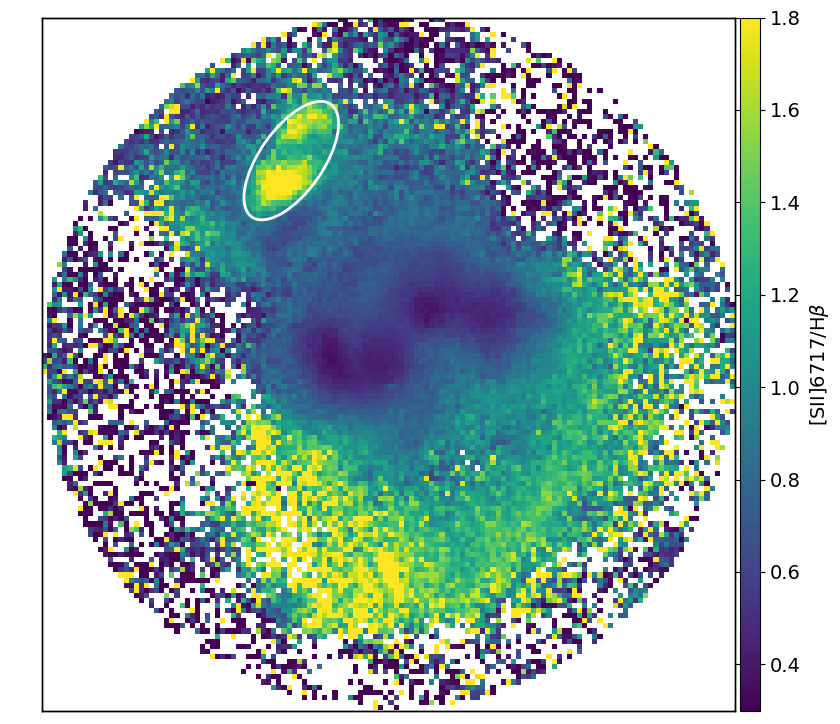}}}
\caption{The ratio of the [SII] lines (left) and [SII]$\lambda$6717/H$\beta$ (right). The white ellipse shows the region used to derive the H$\beta$ flux, i.e. the region where radiative shocks are currently dissipating the jet energy of the microquasar.}
\label{line_ratios}
\end{figure*}


\bsp	
\label{lastpage}
\end{document}